\documentclass[journal,10pt]{IEEEtran}
\usepackage{tabularx}
\usepackage{mathtools}
\newtheorem{lemma}{Lemma}
\usepackage{amsfonts}
\usepackage{algorithm}
\usepackage{algorithmic}
\newcommand{\norm}[1]{\left\lVert#1\right\rVert}
\usepackage{multirow}
\usepackage{stfloats}
\usepackage{graphicx}
\usepackage{color}
\usepackage{indentfirst}

\usepackage{etoolbox}

\newcolumntype{s}{>{\centering\hsize=0.4\hsize\arraybackslash} X}
\newcolumntype{Y}{>{\hsize=1.6\hsize\arraybackslash} X}

\let\mybibitem\bibitem
\renewcommand{\bibitem}[1]{%
\ifstrequal{#1}{ref}
{\color{black}\mybibitem{#1}}
{\color{black}\mybibitem{#1}}%
}

\begin{document}
\graphicspath{ {./img/} }

\title{Optimizing Resource Allocation with High-Reliability Constraint for Multicasting Automotive Messages in 5G NR C-V2X Networks}

\author{
Kuan-Lin Chen, \and 
Wei-Yu Chen, and \and
Ren-Hung Hwang,~\IEEEmembership{senior member,~IEEE}

\thanks{Kuan-Lin Chen, and Ren-Hung Hwang are with the Department of Computer Science and Informational Engineering, National Chung Cheng University, Chiayi, Taiwan. (e-mail: g08410031@ccu.edu.tw; rhhwang@cs.ccu.edu.tw)}
\thanks{Wei-Yu Chen is with the Research Center for Information Technology Innovation (CITI), Academia Sinica, Taipei 115, Taiwan. (e-mail: wei8465@iis.sinica.edu.tw)}}

\markboth{IEEE Transactions on Vehicular Technology,~Vol.~XX, No.~XX, XXX~2021}
{}

\maketitle

\begin{abstract}
Cellular vehicle-to-everything (C-V2X) has been continuously evolving since Release 14 of the 3rd Generation Partnership Project (3GPP) for future autonomous vehicles. Apart from automotive safety, 5G NR further bring new capabilities to C-V2X for autonomous driving, such as real-time local update, and coordinated driving. These capabilities rely on the provision of low latency and high reliability from 5G NR. Among them, a basic demand is broadcasting or multicasting environment update messages, such as cooperative perception data, with high reliability and low latency from a Road Side Unit (RSU) or a base station (BS). In other words, broadcasting multiple types of automotive messages with high reliability and low latency is one of the key issues in 5G NR C-V2X. In this work, we consider how to select Modulation and Coding Scheme (MCS), RSU/BS, Forward Error Correction (FEC) code rate, to maximize the system utility, which is a function of message delivery reliability. We formulate the optimization problem as a nonlinear integer programming problem. Since the optimization problem is NP-hard, we propose an approximation algorithm, referred to as the Hyperbolic Successive Convex Approximation (HSCA) algorithm, which uses the successive convex approximation to find the optimal solution. In our simulations, we compare the performance of HSCA with those of three algorithms respectively, including the baseline algorithm, the heuristic algorithm, and the optimal solution. 
Our simulation results show that HSCA outperforms the baseline and the heuristic algorithms and is very competitive to the optimal solution.

\end{abstract}

\begin{IEEEkeywords}
5G, V2X, multicast, optimal resource allocation, automotive safety\
\end{IEEEkeywords}

\IEEEpeerreviewmaketitle

\section{Introduction}

The 3rd Generation Partnership Project (3GPP) has successively completed the standardization of the 5G Mobile Networks in 2018.
Among the many application scenarios of 5G, V2X application is regarded as the most representative and feasible application of 5G Ultra-Reliable and Low Latency Communications (URLLC).
3GPP has defined service quality requirements of several V2X applications, such as platooning, advanced driving, remote driving, extended sensors \cite{V2X_type1}\cite{V2X_type2}\cite{V2X_type3} in its TS 22.186 V16.2.0 standard specification\cite{3GPP_22.186}.
In order to achieve the above-mentioned applications, high reliability and low latency requirements have become very important.
Compared with 4G Long Term Evolution (LTE), the 5G cellular networks have more radio resources, e.g., the transmission frequency has increased up to 100MHz in the sub-6 GHz frequency range. Thus it can serve more diverse applications and massive devices, such as IoT devices and vehicles. For V2X applications, it can serve more vehicles with high Quality of Service (QoS) requirements, such as high reliability and low latency. In particular, in this work, we focus on investigating how to combine with the multicast service to transmit automotive messages to massive vehicles with high reliability and low latency.

The base transmission unit in 5G is a resource block (RB).
In the definition of 5G, one time slot has 14 symbols in the time domain.
In the frequency domain, one RB contains 12 sub-carriers, and a sub-carrier spacing could be either 15kHz, 30kHz, or 60kHz in the sub-6 GHz spectrum. The slot time for each sub-carrier spacing is 1ms, 0.5ms, and 0.25ms, respectively. 
When a base station (Road Side Unit (RSU) or BS) transmits data to a vehicle, the signal quality received by the vehicle is affected by distance, noise, and interference.
The Signal-to-Interference-plus-Noise Ratio (SINR) plays an important role in deciding the amount of data that can be transmitted in an RB.
Specifically, the amount of data that can be transmitted is determined by Modulation Coding Scheme (MCS), which is related to modulation order and SINR.
The details are defined in 3GPP TS 38.214\cite{3GPP_38.214}.

This work focuses on how to efficiently transmit multiple kinds of automotive messages to massive vehicles with high reliability and low latency. For massive transmission, Single-Cell Point-to-Multipoint (SCPTM) is adopted as the multicast scheme\cite{SC-PTM1}\cite{SC-PTM2}\cite{ref1}. To reduce latency, higher MCS could reduce the number of RBs to transmit automotive messages; however, vehicles with low SINR may not be able to receive the multicast message. That is, reliability may be degraded. To cope with this problem, application-layer Forward Error Correction (FEC) and base station association could help to increase reliability. Thus, the problem is how to achieve the QoS requirements (e.g., bandwidth demand and reliability constraint) of different kinds of automotive messages by joint consideration of base station (or RSU) association, MCS selection, RB allocation, and FEC selection.


In this work, we formulate the multicast problem as an optimization problem that 
maximizes the number of vehicles with satisfied QoS constraints under the constraint of a given limited number of RBs.
Since the transmission reliability is a non-leaner function, we transform the optimization to a dual problem that maximizes the utility while approximating the reliability function by a sigmoid function. Specifically, we propose a "Hyperbolic Successive Convex Approximation (HSCA)" to solve the optimization problem. The main idea of HSCA is based on the successive convex approximation. The proposed HSCA first converts the discontinuous objective function to a continuous function. After the conversion, the gradient of the objective function is then derived and used to guide the direction of the search steps. Our simulation results show that the proposed HSCA algorithm outperforms a baseline algorithm and a heuristic algorithm and can yield very competitive results  compared to the optimal solutions.

The main contributions of this work include:\
\begin{itemize}
    \item To the best of our knowledge, we are the first to present the transmission of multiple types of automotive messages to massive vehicles problem with high reliability and low latency requirements. We formulate the problem as an optimization problem and propose an approximation algorithm that yields very competitive performance to the optimal solution.
    \item To cope with a discontinuous objective function, we apply a successive convex approximation technique and derive a mathematical model to obtain the near-optimal solution iteratively.
    \item To reduce the computational complexity, we also propose a heuristic algorithm that yields competitive solutions as compared to the HSCA algorithm but requires much less computation.
\end{itemize}

The rest of this paper is organized as follows. Section II discusses the related works. In Section III, we introduce the system architecture and system model in the problem formulation. In section IV, we present the design details of the two proposed algorithms, including the heuristic method and the HSCA algorithm. Section V expresses the simulation environment, parameters setting, and simulation results. Finally, conclusions and future works are summarized in Section VI.

\section{Related Work}

In the 5G network, the issue of improving system utilization with the satisfaction of reliability under limited resources is an important research subject, especially for URLLC applications. Improving system utilization by radio resource management has been extensively studied in the literature.
In most of the related studies \cite{Power1}-\cite{Association3}, the common goal is to find the optimal solution for resource allocation to maximize the system utility. In this section, we compare the objective function, constraints, and solutions for these studies. Based on the theme of the study, the related researches are divided into three categories, including BS power control, RB allocation, and user association for the 5G V2X networks.

Recently, some researches \cite{Power1}-\cite{Power4} aimed to maximize the network utility by using BS power control for the V2X communications. Aslani et al. \cite{Power1} proposed an algorithm to optimize the resource and maximize the number of vehicles that can associate with the BS. It adaptively adjusted the BS power to resolve the maximization problem that includes the Doppler effect and power selection. Jameel et al. \cite{Power2} analyzed the mathematical problem by using power control under the constraint of a limited number of available RBs. Their main idea was to apply a Lagrangian method to solve the problem by the power allocation and resource block assignment. Erqing et al. \cite{Power3} developed a strategy to solve a Stackelberg game problem by using power control. The Stackelberg game was adopted to analyze the interaction between the BS and the user equipment (UE) in this study. They also analyzed the Stackelberg model to derive an optimal solution. Regarding the power-efficient issue, the authors of \cite{Power4} proposed a near-optimal ergodic search algorithm to regulate the power consumption at the BS.

Moreover, the investigation of RB allocation for 5G V2X communications has received significant attention recently \cite{RBAllocation1}-\cite{RBAllocation6}. These works either used the entire RB resources while maximizing the network utility or minimized the number of RBs used while satisfying certain QoS constraints. Specifically, Cai et al. \cite{RBAllocation1} proposed a resource scheduling algorithm based on dynamic programming to improve the user's service experience while minimizing the RB usage. He et al. \cite{RBAllocation2} developed a short-term sensing-based resource selection (STS-RS) scheme to maximize the QoS and avoid packet collision while maximizing RB usage. In addition, Le et al. \cite{RBAllocation3} proposed a resource allocation algorithm by applying multi-agent reinforcement learning, which considers the maximum number of available RBs and the number of associated vehicles. This reinforcement learning algorithm achieved an optimal solution by iteratively interacting with the environment. Peng et al. \cite{RBAllocation4} presented an efficient sub-channel allocation to minimize the number of RBs with a power control mechanism. Song et al. \cite{RBAllocation5} proposed a Hungarian algorithm to obtain the appropriate sub-channel assignment and minimize the required number of RBs. This optimization problem is modeled as a mixed binary integer nonlinear programming with channel selection and power control. In the corresponding research \cite{RBAllocation6}, the authors paid attention to construct a location-aware resource allocation (LARA) strategy to minimize the number of RBs and avoid resource collision.

In the 5G V2X communication network, the following literature \cite{Association1}-\cite{Association3} explores the potential of utilizing user association to achieve different objectives. In the traditional method, each UE will associate with the serving BS, which has the best SINR. This may cause some drawbacks, such as traffic load unbalancing. To avoid such drawbacks, Hua et al. \cite{Association1} proposed a BS selection strategy based on a Markov decision policy. They presented a mechanism to predict received signal strength for BS selection and claimed that this strategy is better than the traditional method. On the other hand, Pervej et al. \cite{Association2} developed a distributed single-agent reinforcement learning algorithm to associate UEs to base stations and optimize power allocation. 
Xu et al. \cite{Association3} proposed a weighted-power-based mode selection scheme to associate UEs to base stations. Its superiority lies in the great flexibility to balance the load of the cellular network and improve the coverage probability. By using stochastic geometry, they also derived a model for the base station selection to maximize the coverage
probability.

To sum up, numerous scholars developed various algorithms to achieve different system utilities and resource allocation in the 5G V2X network. We summarize all the references in TABLE \ref{tab:Related works} and divide these studies into three types. Specifically, the algorithms in type 1 apply power control to achieve utility maximization. The schemes of type 2 use different RB allocation strategies to maximize the utility or minimize the usage of RBs. The approaches in type 3 utilize different schemes to associate UEs to base stations. Nevertheless, most of the aforementioned studies lack the consideration of QoS reliability in the constraints. Although some researches \cite{RBAllocation1,RBAllocation2} investigate the utility with reliability, the full consideration of the RBs number limitation and QoS reliability still lacks investigation in the related works. Moreover, according to TABLE \ref{tab:Related works}, some references do not include the mathematical analysis, which is critical to guarantee the effectiveness of the optimal algorithm designs. This paper develops a
theoretical-based approximation algorithm and proposes a heuristic method to resolve the corresponding optimization problem. Both of them significantly improve the system utility while satisfying the QoS reliability constraint and the limit of the number of available RBs of the 5G C-V2X network.

\begin{table*}[t!]
    \centering
    \caption{Comparison of related works}
    \label{tab:Related works}
    \begin{tabular}{|m{0.055\linewidth}<{\centering}|m{0.06\linewidth}<{\centering}|m{0.02\linewidth}<{\centering}|m{0.02\linewidth}<{\centering}|m{0.02\linewidth}<{\centering}|m{0.02\linewidth}<{\centering}|m{0.02\linewidth}<{\centering}|m{0.02\linewidth}<{\centering}|m{0.02\linewidth}<{\centering}|m{0.02\linewidth}<{\centering}|m{0.02\linewidth}<{\centering}|m{0.02\linewidth}<{\centering}|m{0.02\linewidth}<{\centering}|m{0.02\linewidth}<{\centering}|m{0.02\linewidth}<{\centering}|m{0.08\linewidth}<{\centering}|m{0.08\linewidth}<{\centering}|}
    \hline
        \multicolumn{2}{|c|}{Type}& \multicolumn{4}{c|}{1}& \multicolumn{6}{c|}{2}& \multicolumn{3}{c|}{3}& \multicolumn{2}{c|}{This paper} \cr
        \hline
        \multicolumn{2}{|c|}{Algorithms}& \cite{Power1} & \cite{Power2} & \cite{Power3} & \cite{Power4} & \cite{RBAllocation1} & \cite{RBAllocation2} & \cite{RBAllocation3} & \cite{RBAllocation4} & \cite{RBAllocation5} & \cite{RBAllocation6} & \cite{Association1} & \cite{Association2} & \cite{Association3} & Proposed HSCA algorithm & Proposed heuristic algorithm \cr
        \hline
        \multirow{4}{\linewidth}{Objective function} & Maximum system utility & O & O & O & O & O & O & O &&& O & O & O & O & O & O \cr
        \cline{2-17}
        & Minimum number of RBs &&&&&&&& O & O &&&&&&\cr
        \hline
        \multirow{4}{\linewidth}{Constraint} & Number of BS Connection& O & O & O & O & O & O & O && O & O & O & O && O & O \cr
        \cline{2-17}
        & Number of RBs && O &&&&& O & O &&&&& O & O & O \cr
        \cline{2-17}
        & Reliability  &&&&& O & O &&&&&&&& O & O \cr
        \hline
        \multicolumn{2}{|c|}{Mathematical Analysis} & O & O & O && O & O &&&&&& O && O & \cr
        \hline
        \multicolumn{2}{|c|}{Problem Hardness} & NP hard & NP hard &-& NP hard & NP hard & NP hard & NP hard &-& NP hard & NP hard &-& NP hard & NP hard & NP hard & NP hard \cr
        \hline
    \end{tabular}
\end{table*}

\section{Problem Formulation}

\begin{figure}[t!]
    \centering
    \includegraphics[width=0.43\textwidth]{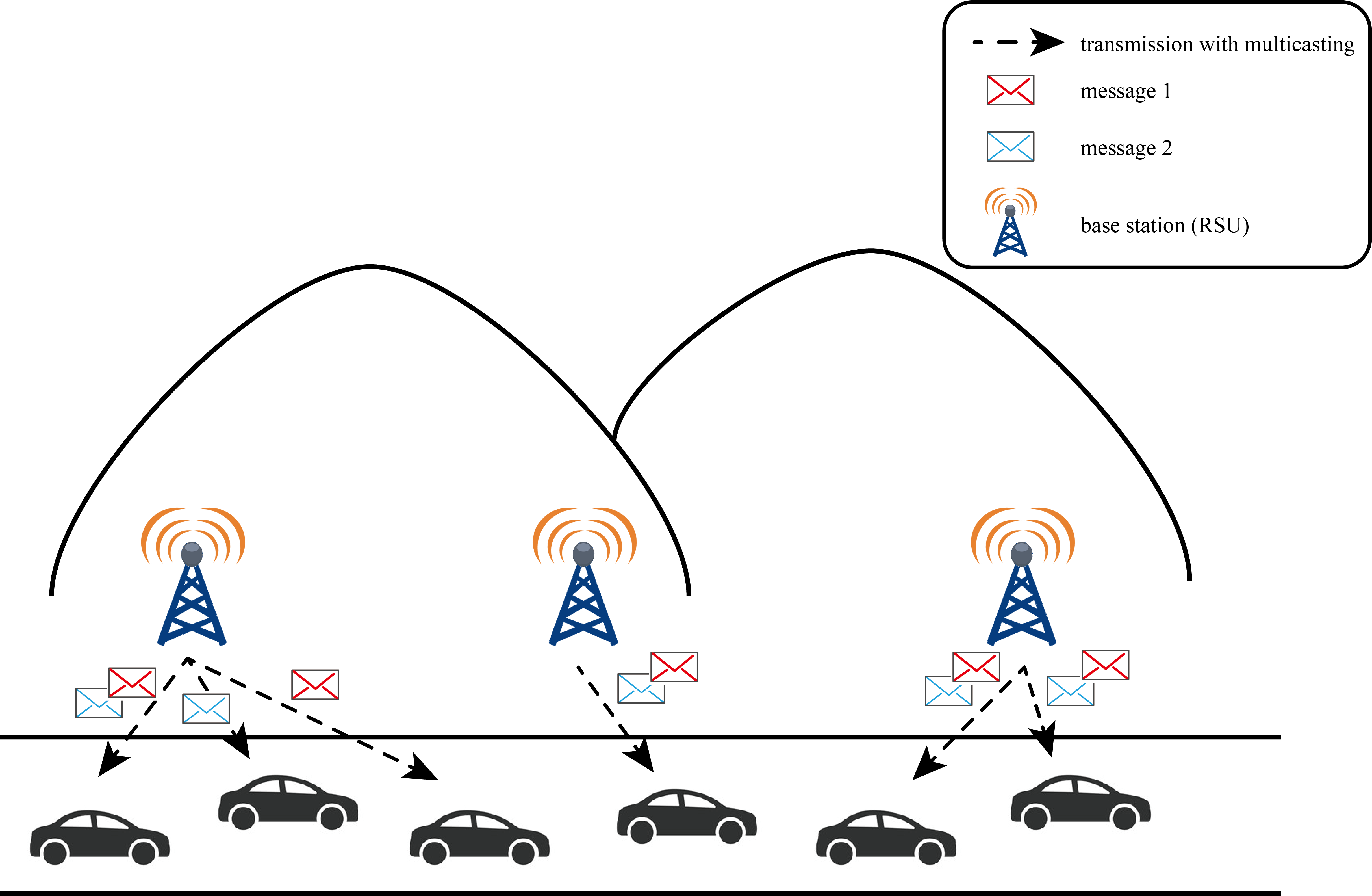}
    \caption{The system architecture of the vehicle transmission message on the highway.}
    \label{fig:system_architecture}
\end{figure}

\subsection{System Architecture}

The system environment of the message transmission is shown in Fig. \ref{fig:system_architecture}.
Multiple BSs are set near the highway. The BS plays the role of RSU, and the type of RSU is picocell. Difference from the traditional BS, picocells are deployed closer to each other than the conventional BS. Each vehicle drives on the highway and needs to receive some messages from the BS. We express the system architecture according to two requirements as follows: First, the transmission needs a low-latency characteristic to support efficient communications. Thus, we apply multicast to transmit the message to all the vehicles.\cite{ref} Second, the transmission requires high reliability, which refers to its QoS requirement. In order to ensure the vehicle can successfully receive messages with high probability, the BS must know the Signal-to-Interference-plus-Noise Ratio (SINR) between the vehicle and the BS and choose Modulation and Coding Scheme (MCS) accordingly. Since different vehicles have different SINR and drive at high speeds, the choice of MCS is critical for achieving high reliability. 
In order to improve the reliability of multicast, controllable mechanisms may include selecting a lower MCS, and adopting the Forward Error Correction (FEC) mechanism at the application layer.

In this paper, we consider a highway scenario with a set of vehicles and BSs (or RSUs). We let $\mathbf{V}$ denote the set of vehicles and $\mathbf{N}$ denote the set of BSs. The BS $n$ has a limited number of RBs per time slot which are reserved for multicast transmission. $M_{n}$ denotes the number of RBs reserved for multicast. Each BS will multicast $\mathbf{K}$ types of messages, and a vehicle $v$ may only be interested in receiving a subset of $\mathbf{K}$ messages.
If vehicle $v$ is associated with the BS $n$ and is interested in receiving type $k$ messages, the allocated number of RBs by BS $n$ for multicasting the messages is denoted by $RB_{n,k}$. 
We let the indicator $w_v^k$ denote whether vehicle $v$ is interested in receiving type $k$ messages or not. Finally, we let the indicator $y_n^v$ denote whether vehicle $v$ is connected to BS $n$ or not. The corresponding definitions of the symbols are summarized in TABLE \ref{tab:notation}.

\begin{table}[t!]
\centering
\caption{Table of Notations}
\label{tab:notation}
\begin{tabularx}{0.48\textwidth}{ |s| >{\arraybackslash}Y|}
\hline
Symbol & Description\cr
\hline
U&System utility\cr
\hline
$\mathbf{K}$&The set of messages, $\mathbf{K}=\{ k_{1},k_{2},k_{3}......,k_{ \mathopen|\mathbf{K}\mathclose|}\}$.\cr
\hline
$\mathbf{V}$&The set of vehicles, $\mathbf{V}=\{ v_{1},v_{2},v_{3}......,v_{ \mathopen|\mathbf{V}\mathclose|}\}$.\cr
\hline
$\mathbf{N}$&The set of BSs, $\mathbf{N}=\{ n_{1},n_{2},n_{3}......,n_{ \mathopen|\mathbf{N}\mathclose|}\}$.\cr
\hline
$y_n^v$&The variable $y_n^v\in\left(0,1\right)(v\in\mathbf{V},\ n\in\mathbf{N})$ denotes whether the vehicle $v$ is served by BS $n$. \cr
\hline
$w_v^k$&The variable $w_v^k\in\left(0,1\right) (v\in\mathbf{V},\ k\in\mathbf{K})$ denotes whether the vehicle $v$ is interested in receiving the $k$ type messages. It is assumed that $w_v^k$ is given initially.\cr
\hline
$S_{n}$&The variable $S_{n}$ denotes the set of vehicles associated with BS $n$. \cr
\hline
$SINR_{n,v}$& $SINR_{n,v}$ is the SINR measured by vehicle $v$ when it is associated with BS $n$.\cr
\hline
$Q_{n,k}$&The variable $Q_{n,k}$ denotes the CQI index (i.e., MCS) for multicasting the message type $k$ selected by the BS $n$.\cr
\hline
$D_k$&The variable $D_k$ denotes the data rate of the message type $k$ which depends on the transmission frequency and the size of the message. It is assumed to be given initially.\cr
\hline
$RB_{n,k}$&The variable $RB_{n,k}$ denotes the number of Resource Blocks allocated by the BS $n$ for multicasting the type $k$ messages.\cr
\hline
$RD_{n,k}$&The variable $RD_{n,k}$ denotes the data rate per RB when the BS $n$ transmits type $k$ message according to the selected $Q_{n,k}$. \cr
\hline
$M_{n}$&The number of reserved RBs by BS $n$ for multicasting the message.\cr
\hline
$F_{n,k}$&The variable $F_{n,k}$ denotes the selected FEC code rate by BS $n$ for the type $k$ message.
\cr
\hline
$p_{n,v,k}$&When the BS $n$ multicasts a type $k$ message  using $Q_{n,k}$, $p_{n,v,k}$ denotes the probability that the vehicle $v$ can successfully receive the message in a RB.\cr
\hline
$PS_{n,v,k}$&The overall probability that the vehicle $v$ in the BS $n$ can successfully receive a type $k$ message, which is calculated using \eqref{equ:PS}.\cr
\hline
$u_{n,v,k}$&The utility of receiving type $k$ messages for the vehicle $v$ associated with the BS $n$.\cr
\hline
$P_k$& The reliability requirements of the type $k$ messages. According to the 3GPP specification for V2X applications, the value is either set to 0.9 or 0.9999.\cr
\hline
$a_k$&The variable $a_k$ denotes the weight of the type $k$ messages which reflects the importance of the messages.\cr
\hline
\end{tabularx}
\end{table}

\subsection{System Model}
Generally, radio resources are scarce in cellular networks.
Although 5G NR has increased the spectrum up to 100MHz in the
sub-6 GHz frequency range, the radio resource is still not enough to satisfy the massive number of devices running with high bandwidth demands, such as AR/VR, 4K videos, and Local Dynamic Map (LDM) for vehicles. Thus, in this paper, we consider the optimization problem that, given a certain amount of radio resources, how to maximize the number of vehicles that can successfully receive different types of V2X messages (e.g., cooperative maneuvers and cooperative perception data) with the desired reliability.
The decision variables of the optimization problem include ${Q}_{n,k}, y_{n}^{v}$, and $F_{n,k}$. The definition of the optimization problem is expressed as follows:\

\begin{equation}
\max_{Q_{n,k},y_{n}^{v},F_{n,k}} U
\label{OPo}
\end{equation}
subject to\
\begin{equation}
\label{equ:constraint1}
M_{n}- \sum_{k=1}^K RB_{n,k} \geq 0 \quad \forall n,\
\end{equation}
\begin{equation}
\label{equ:constraint2}
RB_{n,k} \times RD_{n,k} \times F_{n,k}-D_{k} \geq 0 \quad \forall n,k,\
\end{equation}
\begin{equation}
\label{equ:constraint_y}
\sum_{n=1}^N y_n^v=1 \quad \forall n,\
\end{equation}
\begin{equation}
\label{equ:constraint3}
u_{n,v,k} =
\begin{cases}
a_k \times D_k & PS_{n,v,k} \geq P_k, \\
0 & otherwise.
\end{cases}
\end{equation}
where $U$ is the system utility which is given by:
\begin{equation}
U=\sum_{n=1}^N\sum_{v=1}^V\sum_{k=1}^K u_{n,v,k},
\label{OBJ}
\end{equation}

where $u_{n,v,k}$, the utility of the vehicle $v$ receiving a type $k$ message from the BS $n$, is calculated by \eqref{equ:constraint3}. The detailed definition of each symbol in \eqref{OPo}-\eqref{equ:constraint3} will be described in the following paragraphs and TABLE \ref{tab:notation}.

Constraint \eqref{equ:constraint1} shows that each BS has a limit on the number of RBs that can be used for multicasting. In each BS,  the number of all the required RBs to send $K$ types of messages should not exceed the number of RBs that the BS can provide. $RD_{n,k}$ expresses the required data rate of an RB in a slot time for the $n$th BS and $k$th message. It can be calculated by the selected channel quality indicator (CQI), $Q_{n,k}$, as follows.\

\begin{equation}
\label{equ:RBDR}
    RD_{n,k}=12*14*efficiency(Q_{n,k}).\
\end{equation}

The formulation in \eqref{equ:RBDR} follows the 5G wireless standard. In the frequency domain, one RB consists of 12 sub-carriers, and each sub-carrier has 14 symbols per time slot. $Q_{n,k}$ is the CQI index which corresponds to the specific modulation and code rate. The efficiency of each CQI index is shown in TABLE \ref{tab:MCS_feedback}. In 4G LTE, the CQI index is defined from 0 to 15. Yet, the CQI index is defined from 0 to 31 in the 5G NR standard \cite{3GPP_38.214}. The mapping between the CQI index of 4G, 5G, modulation order, and SINR threshold is also shown in TABLE \ref{tab:MCS_feedback}.
\begin{table}[t!]
    \centering
    \caption{MCS feedback table \cite{3GPP_38.214,MCS_feedback1,MCS_feedback2}}
    \label{tab:MCS_feedback}
    \begin{tabular}{cccccc}
    \hline
    \hline
        4G CQI & 5G CQI & Modulation & Code Rate & SINR & \multirow{2}{*}{efficiency}\\
        index & index & Order & $\times 1024$ & threshold &\\
        \hline
        0 & - & \multicolumn{4}{c}{No transmission}\\
        1 & - & QPSK & 78  & -9.478 & 0.1523\\
        2 & 0 & QPSK & 120  & -6.658 & 0.2344\\
        3 & 2 & QPSK & 193  & -4.098 & 0.3770\\
        4 & 4 & QPSK & 308  & -1.798 & 0.6016\\
        5 & 6 & QPSK & 449  & 0.399 & 0.8770\\
        6 & 8 & QPSK & 602  & 2.424 & 1.1758\\
        7 & 11 & 16QAM & 378  & 4.489 & 1.4766\\
        8 & 13 & 16QAM & 490  & 6.367 & 1.9141\\
        9 & 15 & 16QAM & 616  & 8.456 & 2.4063\\
        10 & 18 & 64QAM & 466  & 10.266 & 2.7305\\
        11 & 20 & 64QAM & 567  & 12.218 & 3.3223\\
        12 & 22 & 64QAM & 666  & 14.122 & 3.9023\\
        13 & 24 & 64QAM & 772  & 15.849 & 4.5234\\
        14 & 26 & 64QAM & 873  & 17.786 & 5.1152\\
        15 & 28 & 64QAM & 948  & 19.809 & 5.5547\\
        \hline
        \hline
    \end{tabular}
\end{table}

On the other hand, FEC is commonly adopted at the application layer to enhance the transmission reliability further.
In this work, block-based rateless FEC is adopted by the BS while multicasting vehicular messages. A code rate of $F_{n,k}$ is expressed as $X/Y$, which indicates that $X$ source blocks are encoded with $Y$ blocks where $Y-X$ blocks are redundant blocks for error correction. A receiver is able to decode these $X$ source blocks if it receives at least $X$ number of blocks out of the $Y$ blocks. Since in 5G networks, the unit of resource allocation is an RB, thus, in this work, the FEC block is in the unit of RBs.
Note that if the transmitted data rate is ${\color{black}RD_{n,k}}$, the effective data rate will be reduced to ${\color{black}RD_{n,k}} \times {\color{black}F_{n,k}}$ when FEC is adopted, as shown in \eqref{equ:constraint2}. \textcolor{black}{Since \eqref{equ:constraint1} implies that $RB_{n,k}$ should be minimized, constraint \eqref{equ:constraint2} can be rewritten as:}
\begin{equation}
\label{equ:RB}
    RB_{n,k}=\left \lceil \frac{D_k}{{\color{black}RD_{n,k}} \times {\color{black}F_{n,k}}} \right \rceil.
\end{equation}

Constraint \eqref{equ:constraint_y} ensures the restriction that one vehicle can only be connected to one BS. If the $v$th vehicle is connected to the $n$th BS, $y_n^v$ will be 1.
Constraint \eqref{equ:constraint3} defines the reliability requirement, which is expressed by a utility function. Specifically, the system can obtain a certain amount of utility only when the probability that 
a vehicle can successfully receive a multicast message is higher than the required reliability. The utility received is proportional to the data rate of the message received and is weighted by a factor that represents the importance of the message. $PS_{n,v,k}$ denotes the probability that the $v$th vehicle in the $n$th BS can successfully receive the $k$th type of message. It can be calculated as follows.
\begin{equation}
\label{equ:PS}
\begin{aligned}
&{PS}_{n,v,k}=w_v^k\times y_n^v\times\\
&\sum_{i=\left\lceil{RB}_{n,k}\times{\color{black}F_{n,k}}\right\rceil}^{{RB}_{n,k}}{\left(\begin{matrix}{RB}_{n,k}\\i\\\end{matrix}\right)(p_{n,v,k})^i{(1-p_{n,v,k})}^{{RB}_{n,k}-i}},
\end{aligned}
\end{equation}
where $p_{n,v,k}$ is the probability that the $v$th vehicle can successfully receive an RB of the $k$th type message, and $i$ is the number of RBs that the vehicle must receive in order to decode the message. 

The calculation of $p_{n,v,k}$ depends on the CQI selected by the BS when multicasting the $k$th type message. TABLE \ref{tab:MCS_feedback} shows the mapping between SINR and CQI when BLER is set to $0.1$ (see \cite{MCS_feedback1,MCS_feedback2}). For example, if the SINR measured by a vehicle is $11.0$, then it will report CQI $10$ to the BS. 
\textcolor{black}{We assume that the channel fading is Rician distribution. The outage probability of a Rician signal received among L Rician/Rayleigh interferers has been derived in the literature. Based on \cite{Rician1}, the outage probability can be expressed as follows.
\begin{equation}
\label{equ:probability}
         \begin{aligned}
        p_{n,v,k}  =  Q \Bigg[ \sqrt{\frac{2LK_I R_I}{b_1+R_I}}; \sqrt{\frac{2K_0 b_1}{b_1+R_I}} \Bigg] \\ +  \exp \Bigg(-\frac{LK_I R_I+K_0 b_1}{b_1+R_I} \Bigg) \\ \times \sum_{m=0}^{L-1} \Bigg(\frac{K_0 R_I}{LK_I b_1}\Bigg)^{m/2} I_m \Bigg( \sqrt{\frac{4LK_I K_0 b_1 R_I}{b_1+R_I}} \Bigg) \\
        \times \Bigg\{ \Bigg(1+ \frac{b_1}{R_I} \Bigg)^{-L} \sum_{m=0}^{L-1}  \begin{pmatrix} L \\ k-m \end{pmatrix} \Bigg(\frac{b_1}{R_I}\Bigg)^k  - \delta_{m0} \Bigg\}
    \end{aligned}
\end{equation}
where $Q$ is the Marcum’s $Q$ function, $L$ is the number of interferers; $m_{i}$, $\sigma_{i}^{2}$, $i=0, ..., L$, are the means and variances of the desired and interfering signals; $K_{I}=\left| m_{I} \right|^{2} / \sigma_{I}^{2}$; $R_{I}$ is the signal-to-interference protection ratio; $K_{0}$ is the Rice factor of the desired signal; $b_{1}= \sigma_{0}^{2} / \sigma_{I}^{2}$; $\delta_{m0}$ is Kronecker delta: $\delta_{m0}=1$ for $m=0$, $\delta_{m0}=1$ for $m \neq 0$. In this work, $R_{I}$ is set to $SINR[Q_{n,k}]$.
}

\section{Algorithm Designs}

In this section, we first proposed a method to initialize vehicle association. This method can allocate all the vehicles to associate with the BS efficiently. After that, we proposed two algorithms to solve our optimization problem. The first algorithm is a theoretical approach that is based on the Successive Convex Approximation (SCA) method to solve the problem. Due to the high computational complexity of SCA, we then propose a heuristic algorithm that iteratively fine-tunes $RB_{n,k}$ while minimizing the number of vehicles that violate their QoS requirement.

\subsection{Vehicle Association}
\label{section:VA}
The optimization problem in \eqref{OPo} mainly consists of two integer decision variables ${\color{black}Q_{n,k}}$, $y_n^v$, and one continuous decision variable ${\color{black}F_{n,k}}$ (assuming rateless coding, e.g., Raptor code or Luby transform (LT) code). 
${\color{black}Q_{n,k}}$ is an integer that ranges from 1 to 15. $y_n^v$ is a binary integer (0 or 1). The objective function and some constraints are nonlinear.
Thus, this optimization problem is a nonlinear mixed-integer programming with discontinuous variables, which is NP-hard and is difficult to solve. 

To simplify the problem, we first fix the variable $y_n^v$ by initializing the association of each vehicle to the BS with the highest SINR. That is, $y_n^v$ is initialized as follows.
 
\begin{equation}
\label{equ:selectBestSINR}
y_n^v = 
\begin{cases}
1 & if\ n=\mathop{\arg \max_{n'}} SINR_{n',v},\\
0 & otherwise.
\end{cases}
\end{equation}
The variable $y_n^v$ is then fine-tuned based on the following heuristic method. After the initial association, the BS $n$ can obtain a group of associated vehicles. It sets ${\color{black}Q_{n,k}}, \forall k$ such that the QoS requirement of the vehicle with the worst SINR can be satisfied. This guarantees that the multicast messages can be received by all the vehicles successfully. For vehicles associated with BS $n$, we then check if the vehicle with the worst SINR could change its association to another BS $n'$ such that the ${\color{black}Q_{n,k}}$ could be increased while $Q_{n',k}$ remains the same. That is, we first find the vehicle $v_{n}^{*}$ that has the minimum SINR in the vehicle group of the BS $n$ as follows.\
\begin{equation}
    v_{n}^{*}= \mathop{\arg \min}_{v \in S_{n}} SINR_{n,v},\
\end{equation}
where $S_{n}$ is a set of vehicle associated with BS $n$. \


We then find a BS $n'$ that meets the following conditions.\
\begin{equation}
\label{equ:ChangeSINRConstraint1}
    Worst\_SINR_{n,k}^*-Worst\_SINR_{n,k}>0,\
\end{equation}
\begin{equation}
\label{equ:ChangeSINRConstraint2}
    Worst\_SINR_{n',k}^*-Worst\_SINR_{n',k}=0,\
\end{equation}
where\
\begin{equation}
\label{equ:afterChangeSINR}
Worst\_SINR_{n,k}^*=\min_{v'\in S_{n}-\{ v_{n}^{*} \} } SINR_{n,v'},\
\end{equation}
\begin{equation}
\label{equ:afterChangeSINR1}
Worst\_SINR_{n',k}^*=\min_{v'\in S_{n'}+\{ v_{n}^{*} \} } SINR_{n',v'},\
\end{equation}
\begin{equation}
\label{equ:afterChangeSINR2}
Worst\_SINR_{n,k}=\min_{v'\in S_{n}} SINR_{n,v'},\
\end{equation}
\begin{equation}
\label{equ:afterChangeSINR3}
Worst\_SINR_{n',k}=\min_{v'\in S_{n'}} SINR_{n',v'}.\
\end{equation}
The constraints \eqref{equ:ChangeSINRConstraint1} and \eqref{equ:ChangeSINRConstraint2} imply that vehicle $v$ can change its serving BS from $n$ to $n'$. We can repeat the calculations from \eqref{equ:selectBestSINR} to \eqref{equ:ChangeSINRConstraint2} for each $n$, until no vehicle can be removed from its current associated BS.

\subsection{Hyperbolic Successive Convex Approximation Algorithm}

By examining the optimization problem, we observe that even if we relax $RB_{n,k}$ to the continuous domain, the utility $u_{n,v,k}$ is still a discontinuous function. Thus, we propose a novel approximation method to approach the original discrete function and verify its effectiveness by simulations. In this algorithm, we do not consider the FEC in the beginning. The optimization problem in \eqref{OPo}-\eqref{equ:constraint3} can be rewritten as a minimizing problem as follows.

\begin{equation}
\min_{{\color{black}Q_{n,k}},y_n^v} -\sum_{n=1}^N\sum_{v=1}^V\sum_{k=1}^K u_{n,v,k}\
\end{equation}
subject to\
\begin{equation}
\label{equ:HSCA_constraint1}
{\color{black}M_n} - \sum_{k=1}^K \left\lceil\frac{D_k}{12\times 14 \times efficiency({\color{black}Q_{n,k}})}\right\rceil \geq0\ \ \forall n,\
\end{equation}
\begin{equation}
\label{equ:HSCA_constraint2}
\sum_{n=1}^N y_n^v=1 \quad \forall n,\
\end{equation}
\begin{equation}
\label{equ:HSCA_constraint3}
u_{n,v,k}=
\begin{cases}
a_k\times D_k & PS_{n,v,k}\geq P_k,\\
0 & otherwise,
\end{cases}
\end{equation}
where the constraint \eqref{equ:HSCA_constraint1} is a combination of \eqref{equ:constraint1} and \eqref{equ:constraint2} according to \eqref{equ:RBDR}-\eqref{equ:RB}. It means that each BS has a limit on the number of RBs for the usage of multicast. The $RB_{n,k}$ can be replaced by \eqref{equ:RBDR} and \eqref{equ:RB} in \eqref{equ:constraint1}. After replacing, we can get the combined constraint \eqref{equ:HSCA_constraint1}. The constraint \eqref{equ:HSCA_constraint2} and \eqref{equ:HSCA_constraint3} remain the same as \eqref{equ:constraint_y} and \eqref{equ:constraint3}.

\begin{figure}[t!]
    \centering
    \includegraphics[width=0.48\textwidth]{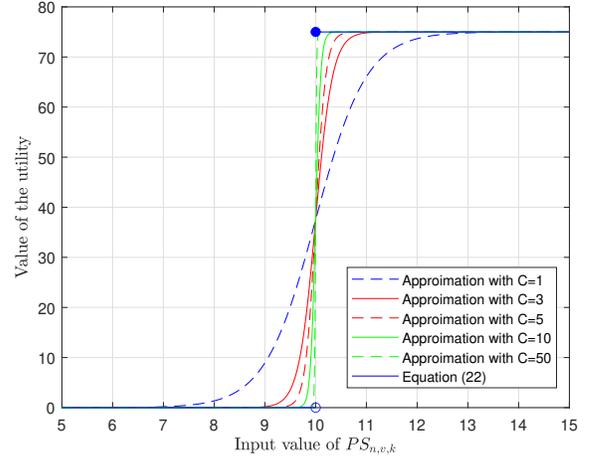}
    \caption{The simple approximation examples of the utility $u_{n,v,k}$ with $a_k=5, D_k=15, P_{k}=10$.}
    \label{fig:continue}
\end{figure}

Second, to transform the objective function into a continuous function, we approximate the utility $u_{n,v,k}$ as a combination of activation functions (i.e., the sigmoid functions). The activation function with different settings of parameter $C$ is illustrated in Fig. \ref{fig:continue}. \textcolor{black}{The motivation for adopting a combination of sigmoid functions is the resulting function is differential (continuous) and could approximate the original utility function well, as shown in Fig. \ref{fig:continue}.}
The equation \eqref{equ:PS} can be rewritten without FEC as follows:
\begin{equation}
    \label{equ:PS_noFEC}
    {PS}_{n,v,k}=w_v^k\times y_n^v\times (p_{n,v,k})^{RB_{n,k}}.
\end{equation}
The approximation of the utility can be expressed as \eqref{equ:HSCA_utility} according to the hyperbolic tangent function and equation \eqref{equ:RBDR}-\eqref{equ:RB}, \eqref{equ:probability}, \eqref{equ:PS_noFEC} without the calculation of FEC. In \eqref{equ:HSCA_utility}, $C>0$ is an arbitrary constant that controls the upward trend of hyperbolic tangent to approach the original discontinuous function. 

\begin{equation}
\label{equ:HSCA_utility}
\begin{aligned}
&u_{n,v,k} \approx a_k D_k \Bigg(2^{-1}\Bigg(tanh\Bigg(C\Bigg(w_v^k y_n^v \times \\
&p_{n,v,k}
^{ \left\lceil\frac{D_k}{12\times 14 \times efficiency({\color{black}Q_{n,k}})}\right\rceil}-P_k\Bigg)\Bigg)+1\Bigg)\Bigg)\
\end{aligned}
\end{equation}

We define the objective function $U_{tot}(\mathbf{x})$ which can be calculated as follows.
\begin{equation}
\label{equ:HSCA_utilityTotal}
U_{tot}(\mathbf{x}) = -\sum_{n=1}^N \sum_{v=1}^V \sum_{k=1}^K u_{n,v,k}
\end{equation}
and the augmented vector is given by:\
\begin{equation}
\begin{aligned}
\mathbf{x}=[&{\color{black}Q_{1,1}},\dots ,{\color{black}Q_{n,k}} \dots,{\color{black}Q_{n,k}},\\
&y_1^1,\dots ,y_n^v \dots,y_N^V]^T.\
\end{aligned}
\end{equation}

The above vector belongs to the solution space $\chi \subset \mathbb{R}^{NK+VN}$. By checking the convexity, we know that a hyperbolic tangent $tanh(\cdot)$ is convex increasing on $\mathbb{R}^-$ and concave increasing on $\mathbb{R}^+$. Since the objective function is a linear combination of the hyperbolic tangent functions, it is trivial that we have to solve a non-convex optimization problem after the above approximation. We have two major difficulties in this optimization problem: First, it has complicated coupling of the two variables. Second, the objective function is a non-convex function for the minimization problem.

Successive convex approximation (SCA) has been shown to be an effective method to resolve various fields of non-convex optimization problems. The main idea of these algorithms is to find the convex surrogate functions at each iteration in order to substitute the intractable non-convex functions. SCA not only gives flexibility in pruning the choice of surrogate functions under efficient aspects but also offers a degree of freedom in the algorithmic design. In the following, we introduce the concept of the proposed SCA algorithm in detail. 
Specifically, we suppose $f(\mathbf{x})$ is a non-convex function. The solution in the $n$th iteration is $\mathbf{x}^n$. According to the original function $f(\mathbf{x})$, the surrogate function is expressed as $\Tilde{f}(\mathbf{x};\mathbf{x}^n)$. Generally, the surrogate function should include the geometric features of $f(\mathbf{x})$ at $\mathbf{x}^n$. Before further investigation, we have to introduce the following useful lemmas \cite{SCA1,SCA2} to select the surrogate function for the non-convex hyperbolic function:\

\begin{lemma}
Let $f(\mathbf{x})$ is any function that the gradient ${\bigtriangledown}f(\mathbf{x}^n)$ of $f(\mathbf{x})$ is Lipschitz continuous with constant $L$ (i.e., the $L$-smooth function). Then for any two solutions $\mathbf{x},\mathbf{x}^n\in X$, we have the following quadratic upper bound property:\
\begin{equation}
f(\mathbf{x}){\leq}f(\mathbf{x}^n)+{\langle}{\bigtriangledown}f(\mathbf{x}^n),\mathbf{x}-\mathbf{x}^n{\rangle}+\frac{L}{2}\|\mathbf{x}-\mathbf{x}^n\|^2\
\end{equation}
\end{lemma}
where $\norm{\cdot}$ denotes the 2-norm of a vector and $\langle\cdot,\cdot\rangle$  is the inner product operator.\\

\begin{lemma}
For an $L_U$-smooth function $U_{tot}\left(\mathbf{x}\right)$ (i.e., the gradient of this function satisfies Lipschitz continuity with constant $L_U$), the surrogate function ${\widetilde{U}}_{tot}\left(\mathbf{x};\mathbf{x}^n\right)$ must be a $L_U$-smooth and convex function, which satisfies the following conditions with given $\mathbf{x},\ {\mathbf{x}}^\prime\in\mathbb{R}^{NK+VN}$ and  $\mathbf{x}^n\in\chi$:\
\begin{equation}
\label{equ:lemma1}
{U}_{tot}\left(\mathbf{x}\right)-{\widetilde{U}}_{tot}\left(\mathbf{x};\mathbf{x}^n\right)\le\frac{{ L}_U}{2}\norm{\mathbf{x}-\mathbf{x}^n}^2+\theta\left(\mathbf{x}^n\right)\
\end{equation}
\begin{equation}
\label{equ:lemma2}
\nabla{\widetilde{U}}_{tot}\left(\mathbf{x}^n;\mathbf{x}^n\right)={\nabla U}_{tot}\left(\mathbf{x}^n\right)\
\end{equation}
\begin{equation}
\label{equ:lemma3}
\norm{\nabla{\widetilde{U}}_{tot}\left(\mathbf{x};\mathbf{x}^n\right)-\nabla{\widetilde{U}}_{tot}\left({\ \mathbf{x}}^\prime;\mathbf{x}^n\right)}\le{L}_U\norm{\mathbf{x}-{\mathbf{x}}^\prime}\
\end{equation}
\end{lemma}
where $\theta\left(\mathbf{x}^n\right)\coloneqq {U}_{tot}\left(\mathbf{x}^n\right)-{\widetilde{U}}_{tot}\left(\mathbf{x}^n;\mathbf{x}^n\right)$ such that the inequality in \eqref{equ:lemma1} becomes equality at $\mathbf{x}=\mathbf{x}^n$.
Condition \eqref{equ:lemma1} means that the block-convex functions should be available to construct effective iterations. The $L_{U}$-smooth function ${U}_{tot}$ can be expressed as ${U}_{tot}=U^s+U^c$, where $U^s$ and $U^c$ are the L-smooth non-convex function and the convex but possibly non-smooth function, respectively. Conditions \eqref{equ:lemma2}-\eqref{equ:lemma3} belong to the first order behavior of $U^s$ and $U^c$ only. According to Lemma 2, ${\widetilde{U}}_{tot}\left(\mathbf{x}^n;\mathbf{x}^n\right)$ is a convex function, and we can replace the update of $\mathbf{x}$ by utilizing ${\widetilde{U}}_{tot}\left(\mathbf{x};\mathbf{x}^t\right)$ instead of the original non-convex function. The quadratic upper bound property in Lemma 1 and the conditions in Lemma 2 suggest that the surrogate function ${\widetilde{U}}_{tot}\left(\mathbf{x};{\mathbf{x}^n}\right)$ could take the following form \cite{SCA1}:
\begin{equation}
{\widetilde{U}}_{tot}\left(\mathbf{x};\mathbf{x}^n\right)={\ U}_{tot}\left(\mathbf{x}^n\right)+\langle{\nabla U}_{tot}\left(\mathbf{x}^n\right),\mathbf{x}\rangle.\
\end{equation}
It is obvious that the above surrogate function is linear with respect to $\mathbf{x}$ and satisfies \eqref{equ:lemma1}, \eqref{equ:lemma2}, and \eqref{equ:lemma3}. According to the above observations, we have the following SCA updated rule:
\begin{equation}
\begin{aligned}
\mathbf{x}^{t+1}&=\mathop{\arg \min}_{\mathbf{x}\in\chi}{{\widetilde{U}}_{tot}\left(\mathbf{x};\mathbf{x}^t\right)}\\
&=\mathop{\arg \min}_{\mathbf{x}\in\chi}{U}_{tot}\left(\mathbf{x}^t\right)+\langle{\nabla U}_{tot}\left(\mathbf{x}^t\right),\mathbf{x}\rangle,
\end{aligned}
\end{equation}
\begin{figure*}
\begin{equation}
\label{equ:gradient}
\begin{aligned}
&{\nabla U}_{tot}\left(\mathbf{x}^t\right)=\left[\begin{matrix}\frac{\partial U_{tot}\left(\mathbf{x}^t\right)}{\partial{Q}}\\\frac{\partial U_{tot}\left(\mathbf{x}^t\right)}{\partial{y}}\\\end{matrix}\right]\\
&=\left[\frac{\partial U_{tot}\left(\mathbf{x}^t\right)}{\partial{Q}_{1,1}}\dots\frac{\partial U_{tot}\left(\mathbf{x}^t\right)}{\partial{Q}_{n,k}}\dots\frac{\partial U_{tot}\left(\mathbf{x}^t\right)}{\partial{Q}_{N,K}}, \frac{\partial U_{tot}\left(\mathbf{x}^t\right)}{\partial y_1^1}\dots\frac{\partial U_{tot}\left(\mathbf{x}^t\right)}{\partial y_n^v}\dots\frac{\partial U_{tot}\left(\mathbf{x}^t\right)}{\partial y_N^V}\right]^T\
\end{aligned}
\end{equation}
\end{figure*}
\begin{figure*}
\begin{equation}
\label{equ:gradientNCQI}
\begin{aligned}
\frac{\partial U_{tot}\left(\mathbf{x}^t\right)}{\partial{Q}_{n,k}}\approx U_{tot}\left(\mathbf{x}^{t-1}\right)-U_{tot}\big([&{Q}_{1,1}\left(t-1\right),\cdots,{Q}_{n,k}\left(t-1\right),\cdots,{Q}_{N,K}\left(t-1\right),\\
&y_1^1\left(t-1\right),\cdots,y_n^v\left(t-1\right),\cdots,y_N^V\left(t-1\right)]^{T}\big)
\end{aligned}
\end{equation}
\end{figure*}
\begin{figure*}
\begin{equation}
\label{equ:gradientY}
\begin{aligned}
\frac{{\partial U}_{tot}\left(\mathbf{x}^t\right)}{\partial y_n^{v}}\approx U_{tot}\left(\mathbf{x}^{t-1}\right)-U_{tot}\big([&{Q}_{1,1}\left(t-1\right),\cdots,{Q}_{n,k}\left(t-1\right),\cdots,{Q}_{N,K}\left(t-1\right),\\
&y_1^1\left(t-1\right),\cdots,y_n^v\left(t\right),\cdots,y_N^V\left(t-1\right)]^{T}\big)
\end{aligned}
\end{equation}
\hrulefill
\vspace{4pt}
\end{figure*}

\noindent
where $t$ is the iteration index of the proposed SCA algorithm. We express the gradient for the SCA updated rule as \eqref{equ:gradient}. In this research, the terms $\frac{\partial U_{tot}\left(\mathbf{x}^t\right)}{\partial{Q}_{n,k}}$ and $\frac{\partial U_{tot}\left(\mathbf{x}^t\right)}{\partial y_n^v}$ are derived by \eqref{equ:gradientNCQI} and \eqref{equ:gradientY} {\footnote{In  \eqref{equ:gradientNCQI} and \eqref{equ:gradientY}, if we calculate the conventional gradient, we must differentiate the entire objective function. However, due to the mathematical nature, the formula after differentiation is complicated and needs a high number of computational steps. Thus, we take the difference of the objective values between the present iteration and previous iteration as the gradient and verify the effectiveness of the proposed HSCA algorithm by the simulations.}}. The updated rules of the proposed SCA can be expressed as:\
\begin{equation}
\label{equ:optX}
\begin{aligned}
\mathbf{x}^{t+1}&=\mathop{\arg \min}_{\mathbf{x}\in\chi}{\ \ {\widetilde{U}}_{tot}\left(\mathbf{x};\mathbf{x}^t\right)}\\
&=\mathop{\arg \min}_{\mathbf{x}\in\chi}{\ {\ U}_{tot}\left(\mathbf{x}^t\right)+\ }\left({\nabla U}_{tot}\left(\mathbf{x}^t\right)\right)^T\mathbf{x.}\
\end{aligned}
\end{equation}
The algorithm only needs to solve a few linear programs and check the feasibility of the constraints \eqref{equ:HSCA_constraint1}-\eqref{equ:HSCA_constraint2} in the iterations. 
The pseudo-code of the proposed hyperbolic successive convex approximation (HSCA) algorithm is expressed in \textbf{Algorithm \ref{alg:HSCA}}.\

\begin{algorithm}[t!]
\caption{Hyperbolic Successive Convex Approximation}
\label{alg:HSCA}
\begin{algorithmic}[1]
\REQUIRE $w_v^k, \alpha_k, D_k, {\color{black}M_n}, P_k,\ \forall n,v,k$
\ENSURE ${Q}_{n,k}, y_n^v, \forall n,v,k$
\STATE Initialize $\left\{{Q}_{n,k},\ \ y_n^v\right\} \forall n,v,k$ as ${x}^0$, where $\mathbf{x}=\left[{Q}_{1,1},\cdots,{\color{black}Q_{n,k}},\ \ y_1^1,\cdots,y_N^V\right]^{T}$
\STATE Calculate the possible $\left\{{Q}_{n,k},\ \ y_n^v\right\}$ as ${x}^1$
\STATE Set the iteration index $t=1$
\FOR{t=1,2,\dots}
\STATE Calculate the gradient ${\nabla U}_{tot}\left(\mathbf{x}^t\right)=\left[\begin{matrix}\frac{\partial U_{tot}\left(\mathbf{x}^t\right)}{\partial{Q}}\\\frac{\partial U_{tot}\left(\mathbf{x}^t\right)}{\partial{y}}\\\end{matrix}\right]$
\STATE Calculate the objective function ${\ U}_{tot}\left(\mathbf{x}^t\right)$
\STATE $\mathbf{x}^{t+1}=\mathop{\arg \min}_{\mathbf{x}\in\chi}{\ \ {\widetilde{U}}_{tot}\left(\mathbf{x};\mathbf{x}^t\right)}=\mathop{\arg \min}_{\mathbf{x}\in\chi}{\ {\ U}_{tot}\left(\mathbf{x}^t\right)+\ }\left({\nabla U}_{tot}\left(\mathbf{x}^t\right)\right)^T\mathbf{x}$
\IF{$\norm{\mathbf{x}^{t+1}-\mathbf{x}^t}^2\le\mathbf{\epsilon}$ or $\mathbf{x}^{t+1}$ violates the constraints \eqref{equ:HSCA_constraint1}-\eqref{equ:HSCA_constraint2}}
\STATE Terminate loop
\ENDIF
\ENDFOR
\STATE Round ${\color{black}Q_{n,k}},\ \forall n,k$ to the integer value according to ${{Q}}^{t}$ in $\mathbf{x}^{t}$
\STATE Round $y_n^v,\forall n,v$ to the binary value according to ${y}^{t}$ in $\mathbf{x}^{t}$
\end{algorithmic}
\end{algorithm}

At line 1 in \textbf{Algorithm 1}, we first initialize all the ${\color{black}Q_{n,k}}$, and $y_n^v$ as $\mathbf{x}^0$. Also, the calculation of $y_n^v$ is obtained by the vehicle association method presented in Section~\ref{section:VA}. We set ${\color{black}Q_{n,k}}$ to 10 as an initial value. The intention is to calculate from high CQI to low CQI in each step with a gradient. At line 2, the initial $y_n^v$ of $\mathbf{x}^1$ is set the same as $\mathbf{x}^0$, and the ${\color{black}Q_{n,k}}$ is set to 9. This process is to calculate the initial gradient by $\mathbf{x}^0$ and $\mathbf{x}^1$. The implementation of HSCA is shown as follows: From line 4 to line 11, we use \eqref{equ:gradient} to calculate the quantities of gradient and utility, and \eqref{equ:optX} is applied to find the corresponding $\mathbf{x}$. If the difference between $\mathbf{x}^{t+1}$ and $\mathbf{x}^{t}$ is less than $\epsilon$, or $\mathbf{x}^{t+1}$ violates the constraints \eqref{equ:HSCA_constraint1}-\eqref{equ:HSCA_constraint2}, we stop the main loop. Otherwise, the algorithm should repeat the instructions from line 4 to line 11. At line 12 and line 13, the whole algorithm terminates after rounding the output value to the nearest integers.

At the end of the algorithm, we can get the approximate solution with ${\color{black}Q_{n,k}}$ and $y_n^v$, and the solution of RBs number can be obtained by \eqref{equ:RB}. However, we conduct the HSCA algorithm without FEC. In other words, there may have extra RBs that can be used. To resolve this issue, we can apply a simple method to allocate the remaining RBs. If an extra RB can improve the highest utility among these messages, we allocate this remaining RB to a specific message type. We loop this method until no remaining RBs can be used.

The complexity of the HSCA algorithm is analyzed as follows. The complexity of computing $U_{tot}\left(\mathbf{x}^t\right)$ and ${\nabla U}_{tot}\left(\mathbf{x}^t\right)$ are $O\left( N \times V \times K \right)$ and $O \left ( N \times (V+K) \right )$, respectively. Since the outer loop repeats until the termination condition is satisfied which we assume to be $T$. Thus, the complexity of HSCA is $O \left ( N \times V \times K \times T \right )$.

\subsection{Heuristic Algorithm}

The computational complexity of the HSCA algorithm may be high for some large scenarios due to the computation of gradients and the steps required for convergence. Thus,
in this subsection, we proposed a heuristic algorithm with low computational complexity. The main idea is to set the ${\color{black}Q_{n,k}}$ to meet the following QoS constraint (taken from \eqref{equ:constraint3}):
\begin{equation}
\label{equ:RBconstraint3}
PS_{n,v,k} \geq P_k \quad \forall n,v,k.
\end{equation}

It then checks if the resulting required number of RBs is larger than the resource constraint (${\color{black}M_n}, \forall n$). If this condition is true, we reduce the $RB_{n,k}$, which will cause the least number of vehicles that violate their QoS constraint. The algorithm will repeat this step until the number of required RBs is less than the resource constraint. The pseudocode of the heuristic algorithm is shown in \textbf{Algorithm 2}.

In \textbf{Algorithm 2}, from line 1 to line 3, each vehicle is associated with the BS by the vehicle association mechanism presented in Section IV-A. From line 4 to line 12, the algorithm uses the worst SINR for all vehicles associated with the same BS to determine the lowest NCQI under the constraint that all vehicles can meet their QoS constraint. Specifically, for each message type $k$, line 7 calculates the worst SINR, denoted by $SINR_{n,k}^\ast$, of all vehicles associated with BS $n$ that are interested in receiving the type $k$ messages. 
Line 8 decreases ${\color{black}Q_{n,k}}$ until all vehicles receiving type $k$ messages can meet the reliability constraint \eqref{equ:RBconstraint3}.
It then derives $RB_{n,k}$ in line 9 and line 10, where ${\color{black}F_{n,k}}$ is set to 1. ${\color{black}F_{n,k}}$ is fine tuned in line 18 when there are residual RBs. In most cases, the sum of the required RBs calculated from lines 4-12 will exceed the available radio resources $({\color{black}M_n})$. From line 13 to line 17, the algorithm tries to reduce the required RBs one by one by minimizing the number of vehicles that would experience QoS violation due to the reduction of RBs. Since the while loop will reduce one RB during each iteration, it is guaranteed that the while loop ends after $(\sum_{n=1}^{N}\sum_{k=1}^{K}{RB_{n,k}}-{M}_{n})$ iterations. The algorithm then returns the solutions of $RB_{n,k}$, ${\color{black}Q_{n,k}}$, and ${\color{black}F_{n,k}}$ at line 19.

 The complexity of the heuristic algorithm is analyzed as follows. The for loop between line 4 and line 12 iterates over all base stations, message types, and vehicles, thus its complexity is $O\left ( N \times V \times K \right )$. For the while loop between line 13 and line 17, its complexity is bounded by $O \left ( {\color{black}M_n} \right )$. For line 18, the running time for the fine tune of FEC usually is negligible, although an upper bound could be given by $O \left ( CQI \times N \times K \right )$. Thus, the complexity of the heuristic algorithm is $O \left ( N \times V \times K + MRB + CQI \times N \times K \right )$ where $MRB=\max\limits_{n}{\color{black}M_n}$. However, in reality, CQI and MRB are much smaller than V, thus we can simply express the complexity of the heuristic algorithm as $O \left ( N \times V \times K \right )$. 

\begin{algorithm}[t!]
\caption{Heuristic resource allocation}
\label{alg:heu}
\begin{algorithmic}[1]
\REQUIRE $D_k, {\color{black}M_n}, P_k, w_v^k\ \forall n,v,k$
\ENSURE ${\color{black}Q_{n,k}}, {\color{black}F_{n,k}}, RB_{n,k}, y_n^v\ \forall n,k$
\FOR{each vehicle $v$}
\STATE Set $y_n^v$ according to the vehicle association mechanism presented in Section IV-A
\ENDFOR
\FOR{each BS $n$}
\FOR{all message type $k$}
\STATE Let $S_{n,v,k}$ be all vehicles associated with BS $n$ \\ $(y_n^v=1)$ and are interested in receiving type $k$ messages $(w_v^k=1)$
\STATE ${SINR}_{n,k}^\ast=\min_{v\in S_{n,v,k}}{{SINR}_{n,v}}$
\STATE Decrease ${Q}_{n,k}$ until ${PS}_{n,v,k}\geq P_k,$  $\forall v,$ \\ where $p_{n,v,k}$ is calculated using \eqref{equ:probability}
with parameters ${SINR}_{n,k}^\ast$ and ${Q}_{n,k}$\\
\STATE ${RD}_{n,k}=12\times14\times efficiency({Q}_{n,k})$
\STATE $RB_{n,k}=\left\lceil\frac{D_k}{{RD}_{n,k}}\right\rceil, {\color{black}F_{n,k}}=1$
\ENDFOR
\ENDFOR
\WHILE{${\color{black}M_n}>\sum_{k=1}^{K}{RB_{n,k}}$}
\STATE Compute $L_{n,k}$=number of vehicles which will violate their QoS when $RB_{n,k}=RB_{n,k}-1$
\STATE $\left(n^\ast,k^\ast\right)={\mathop{\arg \min}_{n,k}{L_{n,k}}}$
\STATE $RB_{n^\ast,k^\ast}=RB_{n^\ast,k^\ast}-1$
\ENDWHILE
\STATE Fine tune ${\color{black}F_{n,k}}$ by increasing ${Q}_{n,k}$ \\
and decreasing $RB_{n,k}$ while the constraint \\
${PS}_{n,v,k}\geq P_k$ still holds
\RETURN $RB_{n,k}, {\color{black}F_{n,k}}, {Q}_{n,k}$
\end{algorithmic}
\end{algorithm}

\section{Simulation} 
In this section, we compare the performance of the proposed two algorithms with the optimal solution (obtained by exhaustive search) and a baseline algorithm. 
The baseline algorithm adopts an intuitive approach which is described as follows. First, vehicles are associated with the BS that has the highest SINR. For each BS, it allocates RBs one by one to different message types by maximizing the increased utility. It repeats the RB allocation until RBs are used up. The baseline algorithm does not adopt the FEC mechanism.

\subsection{System Parameter}
In the simulation, the environment is on a highway. We set up each BS beside the road. Each BS has a circular coverage area with a default radius of 500 meters, and is 1000 meters apart from neighboring BSs. We assume that all the BSs have the same transmit power which is set to 23 dBm. The path loss model belongs to the UMi-Street Canyon scenario in all the simulations \cite{pathloss}. {\color{black}According to the references \cite{NLOS1, NLOS2}, we specify the Rician factor $K=1$. It means the ratio of Line-of-sight to Non-line-of-sight is 1.} The number of reserved RBs per time slot for multicasting varies in each simulation scenario and ranges from 20 to 45. We assume there are five message types, and their detailed configurations are shown in TABLE \ref{tab:messageType} \cite{3GPP_22.186}.
The parameters of the environment are summarized in TABLE \ref{tab:env}.
\begin{table}[t!]
    \centering
    \caption{Configurations of five message types \cite{3GPP_22.186}}
    \label{tab:messageType}
    \begin{tabular}{cccc}
    \hline
        message index & data rate(bit/s) & reliability & weight\\
        \hline
         1 & 100k & 0.9999 & 2\\
         2 & 1000k & 0.9 & 1\\
         3 & 2500k & 0.9 & 1\\
         4 & 50k & 0.99 & 1.5\\
         5 & 2000k & 0.9 & 1\\
         \hline
    \end{tabular}
\end{table}

\begin{table}[t!]
    \centering
    \caption{Default environmental setting}
    \label{tab:env}
    \begin{tabular}{ll}
    \hline
         Parameter & Value \\
         \hline
         Cell Radius & 500m \\
         Small-scale fading & Rician fading\\
         RSU & Pico cell\\
         BS Power & 23dBm\\
         Transmit antenna power gain & 1dBi\\
         Center frequency & 5.9GHz\\
         Bandwidth & 20MHz\\
         Path Loss & $32.4+20Log_{10}(f_c)+$ \\
         & $31.9Log_{10}(d)$ \cite{pathloss}\\
         Lognormal shadowing & 8.2 dB standard deviation\\
         Ratio of NLOS and LOS & 1:1\cite{NLOS1,NLOS2}\\
         Noise power density & -174 dBm/Hz\\
         Antenna number at vehicle & 2\\
         Antenna number at BS & 4\\
         RB size & 12 sub-carrier, 1 ms time slot \cite{3GPP_38.214}\\
         Sub-carrier spacing & 15KHz\\
         Vehicle speed & 90$\sim$110 km/hr\\
         \hline
    \end{tabular}
\end{table}

\begin{figure}[t!]
    \centering
    \includegraphics[width=0.48\textwidth]{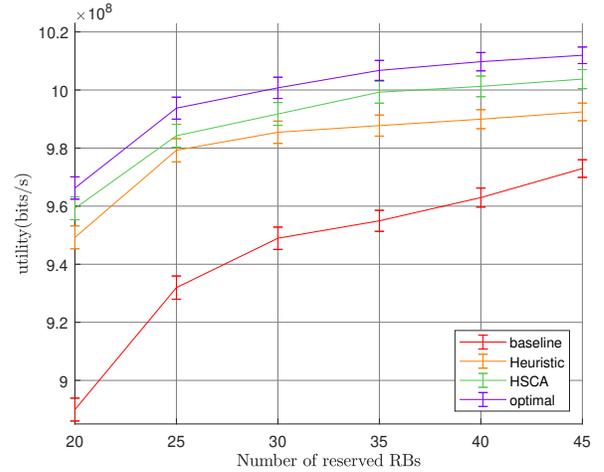}
    \caption{The average system utility under various number of reserved RBs with 250 vehicles and 5 message types.}
    \label{fig:5message}
\end{figure}

\begin{figure*}[t!]
    \centering
    \includegraphics[width=\textwidth]{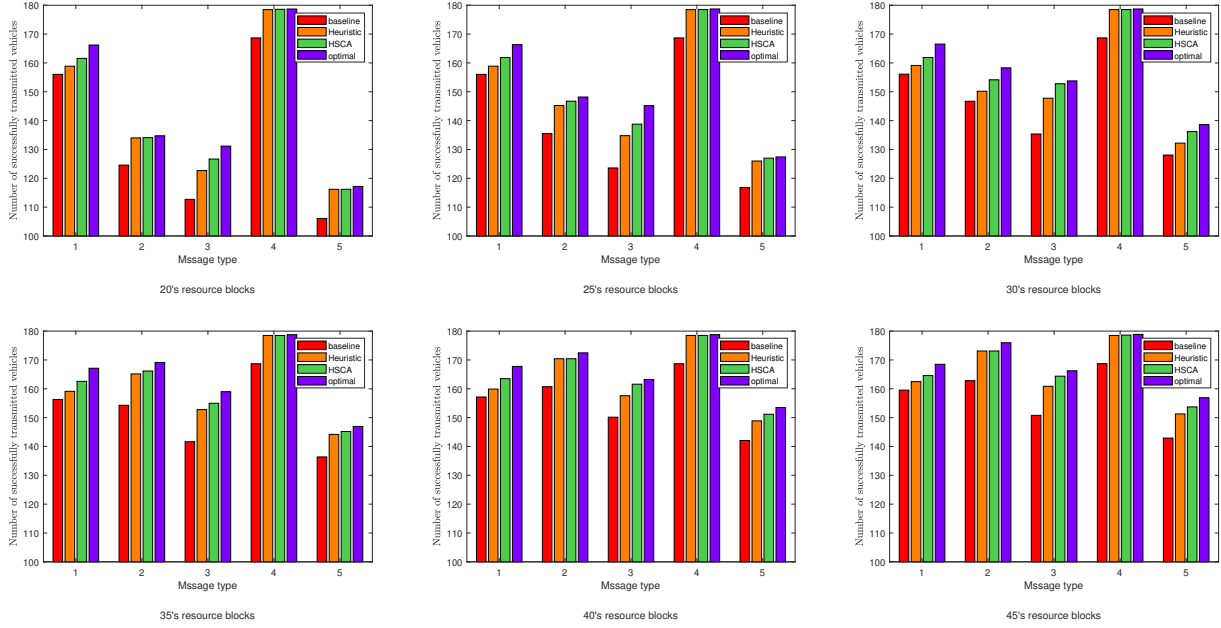}
    \caption{Number of successfully transmitted vehicles under various number of reserved RBs with 250 vehicles and 5 message types.}
    \label{fig:throughput}
\end{figure*}

\subsection{Performance Evaluation}
Unless otherwise stated, we assume that there are up to 250 vehicles on the highway. By default, each vehicle will receive five types of messages from the BS. Each simulation is run for 1000 time slots. In each figure, the vertical bar at each simulated point indicates the $95\%$ confidence interval.

Fig. \ref{fig:5message} shows the system utility of the proposed heuristic and HSCA algorithms under the different number of reserved RBs. The number of vehicles is set to 250. We can observe that both the proposed algorithms have significantly superior performance than the baseline algorithm, and HSCA yields the best utility among these three algorithms. Furthermore, the utility yielded by HSCA is very close to the optimal solution.  When the RBs' number is 20, the difference between the proposed two algorithms is not very significant. But as the number of reserved RBs increases, the performance of HSCA becomes more significantly better than that of the heuristic algorithm. In addition, Fig. \ref{fig:throughput} shows the throughput of different message type. We define the throughput as the average number of vehicles that have successfully received the multicast message transmitted from the RSUs with the designated reliability requirement. We can observe that the HSCA algorithm and the heuristic algorithm yield competitive throughput compared to the optimal solution while they perform much better than the baseline algorithm. We also observe that as the number of RB increases, the throughput also increases, as the RSUs have more resources to multicast V2X messages. Since the design of the utility metric takes reliability into account by giving higher weights to higher reliability, we also observe that the throughput of message type 1 and message type 4 is higher than the other three message types. Nonetheless,
the performance shown in Fig. \ref{fig:throughput} is consistent with that in Fig. \ref{fig:5message} which implies that 
the utility metric still acts as a good indicator of throughput.

\begin{figure}[t!]
    \centering
    \includegraphics[width=0.48\textwidth]{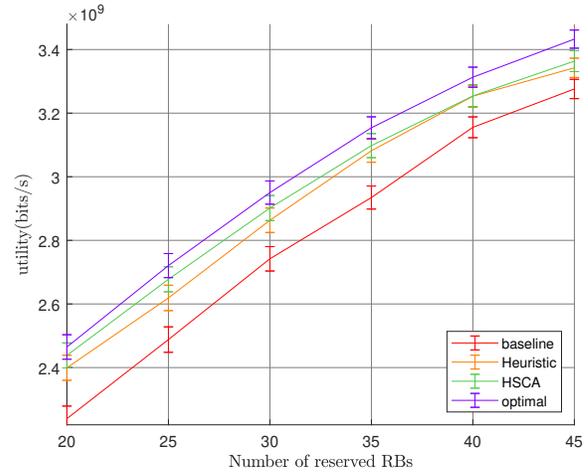}
    \caption{The average system utility versus different limitations of RBs with 1000 vehicles, 5 message types.}
    \label{fig:1000vehicle}
\end{figure}

Fig. \ref{fig:1000vehicle} compares the utility of different algorithms when the number of vehicles is set to 1000. As the number of vehicles increases, the number of vehicles with low SINR also increases. Thus more RBs are required to satisfy these vehicles with low SINR. From Fig. \ref{fig:1000vehicle}, we can observe that the system utility increases more linearly as compared to Fig. \ref{fig:5message}, and the differences between different algorithms become less. However, the trend of the performances of the three algorithms remains the same. The HSCA algorithm still yields the best performance and is very competitive to the optimal solution. But the performance of the heuristic algorithm becomes very close to that of HSCA.

\begin{figure}[t!]
    \centering
    \includegraphics[width=0.48\textwidth]{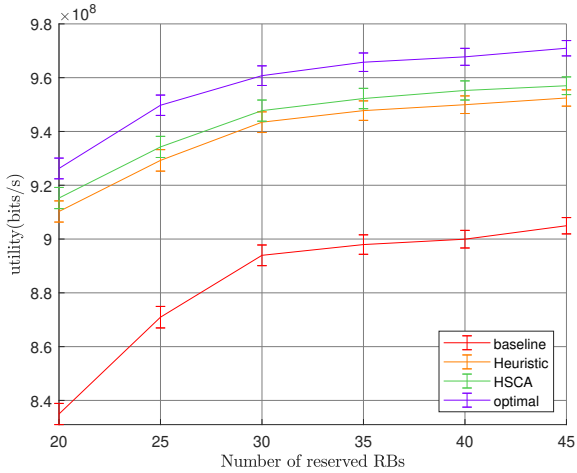}
    \caption{The average system utility versus different limitations of RBs with 250 vehicles, 5 message types, and the BSs are deployed according to the binomial point process.}
    \label{fig:binomial}
\end{figure}

In the next simulation scenario, the distance between two BS is not fixed to 1000 meters, but the BSs are deployed according to the Poisson Point Process (which becomes a Binomial Point Process as we fixed the number of BSs to 5). We generate 100 deployment scenarios, and for each scenario, we run the simulation for 1000 time slots. Fig. \ref{fig:binomial} shows the system utility generated by four algorithms. The results are similar to the case of fixed deployment of BSs, except that the system utility slightly degrades due to the uneven distribution of BSs. In addition, the difference between the HSCA and the optimal solution becomes larger, while the difference between the HSCA and the heuristic algorithm becomes smaller.


\begin{figure}[t!]
    \centering
    \includegraphics[width=0.48\textwidth]{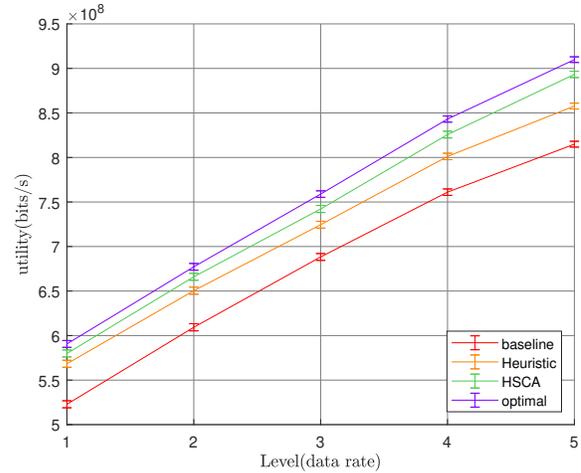}
    \caption{The average system utility versus different levels of data rate with 250 vehicles, 5 message types, and different data rates.}
    \label{fig:5level}
\end{figure}

In the next simulation scenario, we examine the effect of the data rate of message types on the performance of different algorithms. TABLE \ref{tab:5level} shows five levels of data rates for five types of messages. Fig. \ref{fig:5level} shows the system utility yielded by the four algorithms. Since the date rate increases from level 1 to level 5 for the first three data types, the utilities of four algorithms also increase from level 1 to level 5. The relative performance among the four algorithms remains the same, while the utility yielded by HSCA is very close  to that of the optimal solution.

\begin{table}[t!]
    \centering
    \caption{The data rate(bits/s) of the five message types which are divided into 5 levels}
    \label{tab:5level}
    \begin{tabular}{cccccc}
    \hline
         message index & Level 1 & Level 2 & Level 3 & Level 4 & Level 5 \\
         \hline
         1 & 50k & 340k & 630k & 920k &1200k\\
         2 & 300k & 325k & 350k & 375k & 400k\\
         3 & 50k & 112k & 175k & 240k &300k\\
         4 & 1600k & 1600k & 1600k & 1600k & 1600k\\
         5 & 2000k & 2000k & 2000k & 2000k & 2000k\\
         \hline
    \end{tabular}
\end{table}

\begin{figure}[t!]
    \centering
    \includegraphics[width=0.48\textwidth]{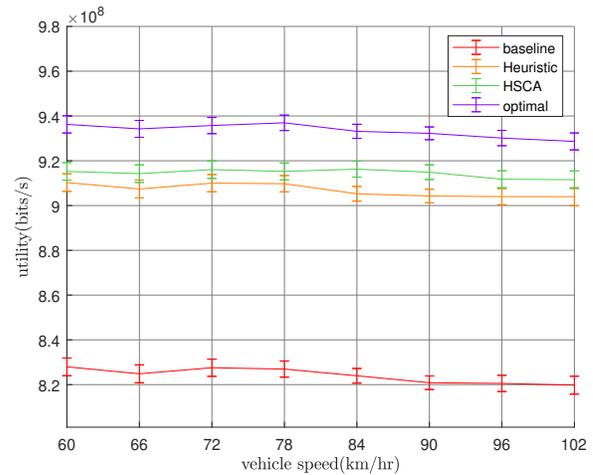}
    \caption{The average system utility versus different vehicle speeds with 250 vehicles, 5 message types, and average speeds of vehicles range from 60 to 102 km/hr}
    \label{fig:speed}
\end{figure}

Fig. \ref{fig:speed} examines the system utility under different vehicle speeds. The average speed of the vehicle is ranged from 60 to 102 km/hr. For a given average speed $V\_avg$, the speed of each vehicle is randomly generated within the range of [$V\_avg$-10,$V\_avg$+10] km/hr. In general, as we can observe from Fig. \ref{fig:speed}, the system utility is not affected by the speeds of vehicles, except that when the vehicle's speed is very fast (e.g., higher than 90 km/hr), the handoff between two BSs and the change of vehicles' SINR becomes more frequent. Thus, the utility of each algorithm is slightly degraded as the average speed of vehicles is larger than 90 km/hr.


\begin{figure}[t!]
    \centering
    \includegraphics[width=0.48\textwidth]{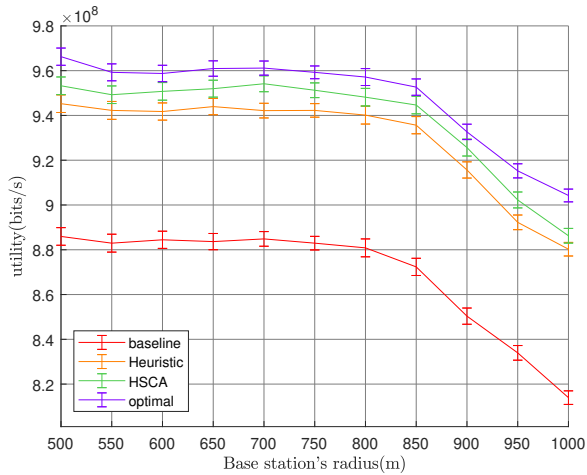}
    \caption{The average system utility versus different levels of BS's radius with 250 vehicles, 5 message types, and cell radius ranges from 500m to 1000m.}
    \label{fig:radius}
\end{figure}

In the next simulation scenario, we explore the effect of the cell radius of BSs on the system utility in Fig. \ref{fig:radius}. Note that BSs are deployed such that the distance between two BSs is set to the cell radius. As a consequence, vehicles at the cell edge may experience lower SINR as the cell radius increases since the transmission power of BS is kept at 23 dBm. As we can observe from Fig. \ref{fig:radius}, the relative performance among the four algorithms remains the same. However, we notice that the system utility degrades dramatically when the cell radius is larger than 800 meters, mainly due to the degradation of SINR of cell edge vehicles.


\begin{figure}[t!]
    \centering
    \includegraphics[width=0.48\textwidth]{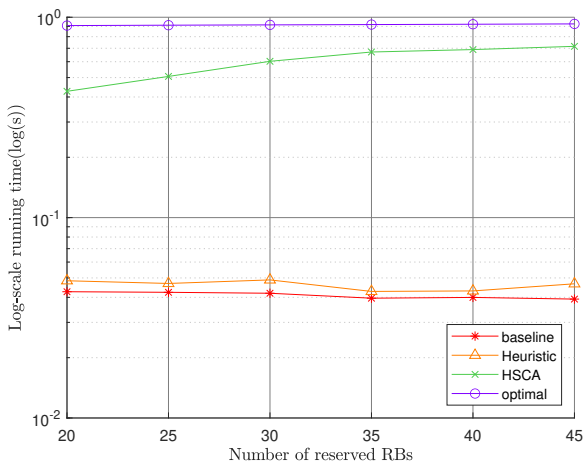}
    \caption{The running time under various number of reserved RBs with 250 vehicles and 5 message types.}
    \label{fig:runningTime}
\end{figure}

Finally, in Fig. \ref{fig:runningTime}, we compare the running time of different algorithms. We conducted the simulations using MATLAB 2020b, which ran on a computer with a 16-core  CPU, 32G RAM, and Ubuntu 18.04 operation system. We can observe that the running times for the HSCA and optimal algorithms are much higher than that of the baseline and heuristic algorithms which justifies why, in addition to the HSCA algorithm, we also propose the heuristic algorithm. Although HSCA runs faster than the optimal algorithm, however, the calculation of the gradient matrix consumes very high computation time. Besides, as the number of reserved RBs increases, it takes more iteration to converge; thus, its running time also increases. By comparing the HSCA and the baseline algorithm, we can observe a trade-off between performance and running time.

\section{Conclusions and future work}
This paper investigated the resource allocation optimization problem to boost the system utility under several constraints of QoS reliability in a 5G NR V2X environment. We first defined the optimization problem and then proposed two algorithms, i.e., the HSCA algorithm and heuristic method, to maximize the system utility. In particular, the HSCA is designed based on theoretical analysis and successive convex approximation framework. The proposed algorithms provide effective solutions to satisfy various reliability demands of vehicle safety messages while maximizing the system utility. In the simulation results, we verify the effectiveness of the proposed algorithms by comparing the proposed algorithms with the optimal solution. Our simulation results show that both the proposed HSCA and heuristic algorithms perform better than the baseline algorithm and yield very close utility to the optimal solution. 

As V2V communication becomes more and more important in 5G NR V2X networks, it is very promising to further improve the transmission reliability of vehicular safety messages. Thus, we are investigating the issue of how to combine multicasting and V2V communication techniques to further improve the reliability of message transmission and system utility. 
\textcolor{black}{On the other hand, solving the optimization problem using HSCA is time-consuming. Recently, deep learning has shown its promise in solving the optimization problem in 6G networks \cite{6GDL}. Thus, applying deep learning to our optimization problem is worthy of future investigation.}

\end{document}